\documentclass[aps,twocolumn,epsfig,prl]{revtex4}

\usepackage{amsmath}
\usepackage{graphicx}

\begin{document}

\title{Berry's Phase for Standing Wave Near Graphene Edge}

\author{Ken-ichi Sasaki}
\email[Email address: ]{SASAKI.Kenichi@nims.go.jp}
\affiliation{International Center for Materials Nanoarchitectonics, 
National Institute for Materials Science,
Namiki, Tsukuba 305-0044, Japan}

\author{Katsunori Wakabayashi}
\affiliation{International Center for Materials Nanoarchitectonics, 
National Institute for Materials Science,
Namiki, Tsukuba 305-0044, Japan}
\affiliation{PRESTO, Japan Science and Technology Agency,
Kawaguchi 332-0012, Japan}

\author{Toshiaki Enoki}
\affiliation{Department of Chemistry, Tokyo Institute of Technology,
Ookayama, Meguro-ku, Tokyo 152-8551, Japan}

\date{\today}

\begin{abstract}
 Standing waves near the zigzag and armchair edges, and their Berry's phases
 are investigated.
 It is suggested that the Berry's phase for the standing wave near the
 zigzag edge is trivial, while that near the armchair edge is non-trivial. 
 A non-trivial Berry's phase implies 
 the presence of a singularity in parameter space. 
 We have confirmed that the Dirac singularity is absent (present) in the
 parameter space for the standing wave near the zigzag (armchair) edge.
 The absence of the Dirac singularity 
 has a direct consequence in the local density of states near the zigzag edge.
 The transport properties of graphene nanoribbons observed by recent
 numerical simulations and experiments are discussed from the point of
 view of the Berry's phases for the standing waves. 
\end{abstract}


\maketitle

The electronic property of a graphene nanoribbon
differs greatly from that of a carbon nanotube.
A metallic carbon nanotube exhibits a high mobility, 
while a graphene nanoribbon shows a transport
gap.~\cite{chen07,stampfer09,han10,gallagher10}
The high mobility observed in metallic carbon nanotubes
indicates that the scatterers are not effective 
in producing backward scattering.
There should be a mechanism 
which suppresses the backward scattering
in metallic carbon nanotubes.

\begin{figure}[htbp]
 \begin{center}
  \includegraphics[scale=0.4]{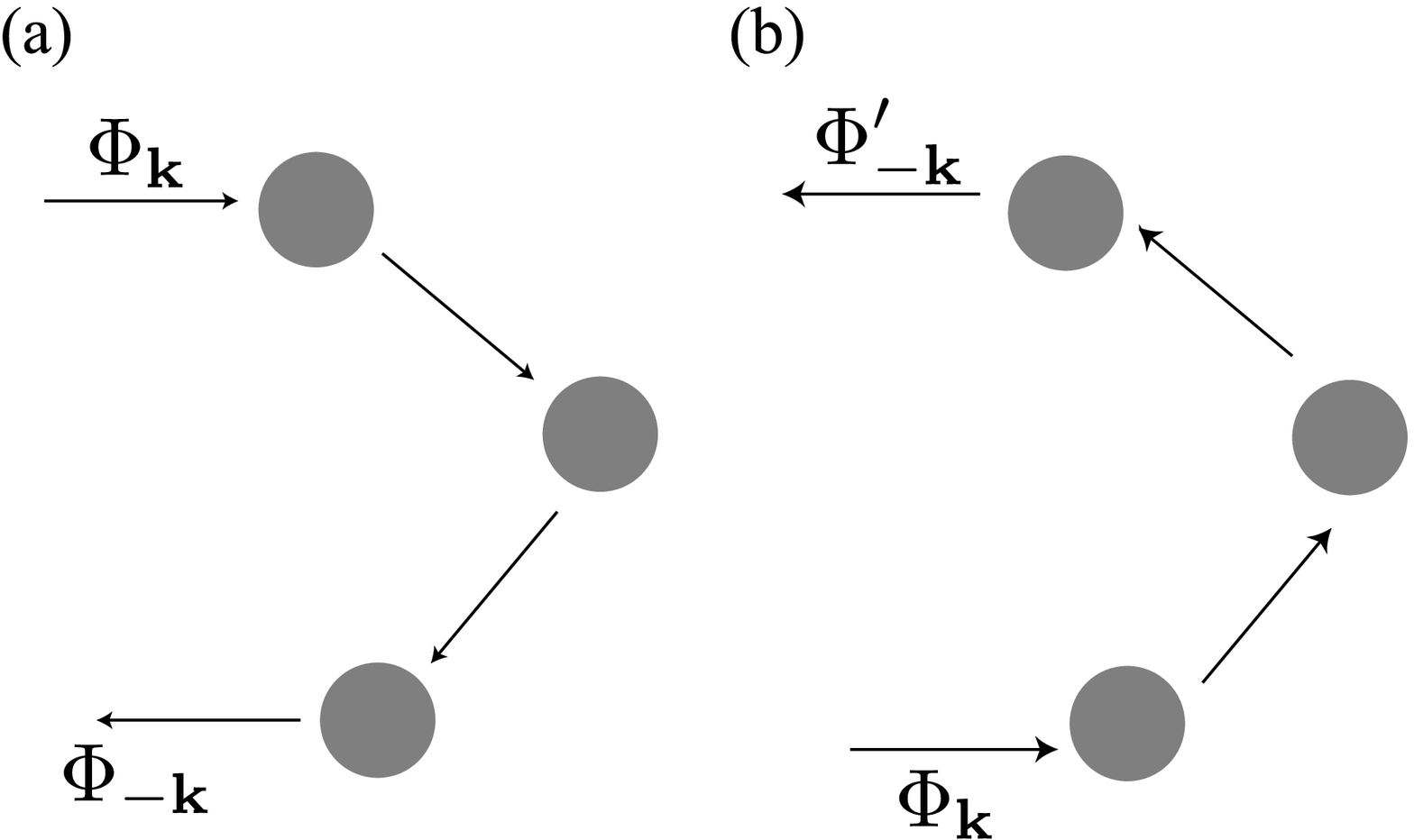}
 \end{center}
 \caption{The finial state of the scattering path shown in (a),
 $\Phi_{\bf -k}$, combines with the final state of the scattering path
 shown in (b), $\Phi'_{\bf -k}$, to suppress the total backscattering
 amplitude, $\Phi_{\bf -k}+\Phi'_{\bf -k}$, due to the Berry's phase. 
 The two scattering paths,
 (a) $\Phi_{\bf k}\to \cdots \to \Phi_{\bf -k}$ and 
 (b) $\Phi_{\bf k}\to \cdots \to \Phi'_{\bf -k}$, 
 are related with each other by ``time-reversal''. 
 }
 \label{fig:abs}
\end{figure}

Suppose that an electron with momentum ${\bf k}$ 
is coming into the impurities which are represented by 
the circles in Fig.~\ref{fig:abs}(a). 
The electron is scattered by the impurities and 
changes its momentum direction. 
Let us consider the probability amplitude that 
the electron is scattered in the backward direction, 
as shown in Fig.~\ref{fig:abs}(a). 
In this case, the final wave function, $\Phi_{\bf -k}$, 
is given by rotating the wave vector of the initial state,
$\Phi_{\bf k}$, by $-\pi$,
so that we have the relationship between the initial state
and final state as
\begin{align}
 \Phi_{\bf -k}=U(-\pi)\Phi_{\bf k},
 \label{eq:path1}
\end{align}
where $U(\theta)$ is a rotational operator with angle $\theta$.
Note that this particular path shown in Fig.~\ref{fig:abs}(a)
is not the unique path that an electron can follow. 
There is an another path that an electron can follow, 
which we denote it by the lines in Fig.~\ref{fig:abs}(b). 
The new path relates to the original path through the ``time reversal''. 
We denote the final state in this ``time reversal'' path by $\Phi'$. 
Because the final state is given by rotating the wave vector by $+\pi$, 
we have the relationship, 
\begin{align}
 \Phi'_{\bf -k}=U(+\pi)\Phi_{\bf k},
 \label{eq:path2}
\end{align}
between the initial state and the final state. 
By eliminating the wave function of the initial state from Eqs.~(\ref{eq:path1})
and (\ref{eq:path2}), we get
\begin{align}
 \Phi'_{\bf -k}=U(2\pi)\Phi_{\bf -k}.
\end{align}
Now, the total backward scattering amplitude 
is given by the sum of $\Phi_{\bf -k}$ and $\Phi'_{\bf -k}$ as
$\Phi_{\rm bs}=\Phi'_{\bf -k}+\Phi_{\bf -k}=\left[ 1+U(2\pi)\right] \Phi_{\bf -k}$.
Because the wave function 
gets an extra phase shift of $-\pi$ 
(called the Berry's phase~\cite{berry84,simon83}) through 
a rotation of the wave vector around the Dirac point,
that is, $U(2\pi)$ is equivalent to $-1$ ($U(2\pi)=e^{-i\pi}$),
a ``time-reversal'' pair of backward scattered waves 
cancels with each other, i.e., $\Phi_{\rm bs}=0$~\cite{ando98}.

This, absence of backward scattering mechanism
provides us with a simple solution explaining the high mobility, 
and the existence of a singularity at the Dirac point 
is essential to the nontrivial phase shift of $\pi$.
However, it is not obvious 
whether the wave functions realizing in graphene nanoribbons
can acquire a nontrivial Berry's phase or not. 
The eigen state near the edge is 
the standing wave resulting from the interference between 
an incident wave and the edge reflected wave. 
If the Berry's phase for the standing wave is trivial, 
then the scatters, such as a potential disorder created by charge
impurities, may give rise to backward scattering.
As a result, the mobility of a graphene nanoribbon decreases
considerably than that of a carbon
nanotube.~\cite{chen07,stampfer09,han10,gallagher10}  
In the present paper,
we study the Berry's phases of the standing waves near the zigzag and
armchair edges, and their effects on the local density of states
and the transport properties of graphene nanoribbons.

\begin{figure}[htbp]
 \begin{center}
  \includegraphics[scale=0.4]{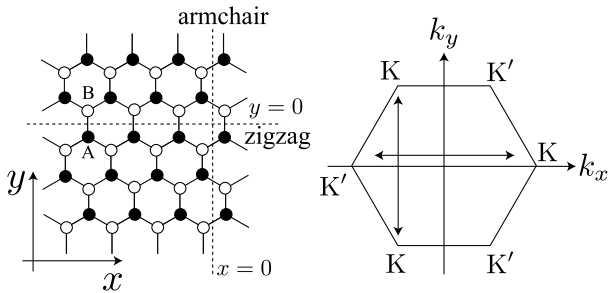}
 \end{center}
 \caption{
 A hexagonal unit cell of graphene consists of 
 {\rm A} (solid circle) and {\rm B} (open circle) atoms. 
 The $xy$ coordinate system is fixed as shown.
 (right) The hexagon represents the first BZ of graphene, 
 and the corners of the hexagon are K and K$'$ points.
 }
 \label{fig:unit}
\end{figure}

In Fig.~\ref{fig:unit},
we consider the zigzag edge parallel to the $x$-axis, 
by which translational symmetry along the $y$-axis is broken.
Thus, the incident state with wave vector $(k_x,k_y)$ 
is elastically scattered by the zigzag edge, 
and the wave vector of the reflected state becomes $(k_x,-k_y)$.
By contrast,
the armchair edge parallel to the $y$-axis 
breaks translational symmetry along the $x$-axis, 
so that the wave vector of the reflected state is $(-k_x,k_y)$.
Since the Brillouin zone (BZ) 
is given by rotating the hexagonal lattice by 90$^\circ$,
one can see in Fig.~\ref{fig:unit} that 
for the incident state near the K point, 
the reflected state by the zigzag edge
is also near the K point, while
the reflected state by the armchair edge is near the K$'$ point. 
Hence, the scattering by the zigzag edge is intravalley scattering, 
while that by the armchair edge is intervalley scattering.

To begin with, 
let us consider the scattering problem for the zigzag edge.
Since the zigzag edge is not the source of intervalley scattering,
we focus on only the electrons near the K point.
The incident and reflected waves 
are represented by the Bloch states.
The Bloch state in the conduction energy band 
is written as
\begin{align}
 \Phi_{{\rm K},{\bf k}}^c({\bf r}) = 
 \frac{e^{i{\bf k}\cdot {\bf r}}}{\sqrt{V}} 
 \phi_{{\rm K},{\bf k}}^c, \ \ 
 \phi_{{\rm K},{\bf k}}^c = \frac{1}{\sqrt{2}}
 \begin{pmatrix}
  e^{-i\theta({\bf k})} \cr 1
 \end{pmatrix},
\end{align}
where $V$ is the area of sample,
${\bf k}$ is the momentum measured from the K point,
and $\theta({\bf k})$ is the polar angle between 
the vector ${\bf k}$ and the $k_x$-axis. 
Note that the Bloch function 
acquires a nontrivial Berry's phase of $-\pi$ 
by a rotation of the wave vector around the K point as
\begin{align}
 \int_0^{2\pi} 
 \langle \phi_{{\rm K},{\bf k}}^c|
 \frac{1}{i}\frac{\partial}{\partial \theta}|
 \phi_{{\rm K},{\bf k}}^c \rangle d\theta 
 = -\pi.
 \label{eq:berry-}
\end{align}

The standing wave modes arise from the interference 
between the incident waves and the reflected waves.
Let the wave vector of the incident wave be ${\bf k}=(k_x,k_y)$.
Then the vector of the elastically reflected wave 
is given by ${\bf k'}=(k_x,-k_y)$, 
and the standing wave near the zigzag edge is written as
\begin{align}
 \Psi_{{\rm K},{\bf k}}^c({\bf r}) = 
 \Phi_{{\rm K},{\bf k}}^c({\bf r})+ e^{iz} \Phi_{{\rm K},{\bf
 k'}}^c({\bf r}),
 \label{eq:zigwfsum}
\end{align}
where the phase $e^{iz}$ should be determined
in such a manner that $\Psi_{{\rm K},{\bf k}}^c({\bf r})$
[$={}^t(\Psi^{c}_{{\rm K},{\bf k},{\rm A}}({\bf r}),\Psi^{c}_{{\rm K},{\bf k},{\rm B}}({\bf r}))$]
satisfies the boundary condition~\cite{berry87}.
We take the following boundary condition for the zigzag edge~\cite{brey06},
\begin{align}
 \lim_{y\to 0}\Psi^{c}_{{\rm K},{\bf k},{\rm B}}({\bf r}) = 0,
 \label{eq:phiB0}
\end{align}
which describes a situation in which
the B-atoms located slightly above the horizontal dashed line at $y=0$
in Fig.~\ref{fig:unit} are separated from the lower semi-infinite
graphene for $y<0$. 
The boundary condition of Eq.~(\ref{eq:phiB0})
is satisfied when $e^{iz} = -1$, and 
the standing wave is written as
\begin{align}
 \Psi^{c}_{{\rm K},{\bf k}}({\bf r}) = \frac{e^{ik_x x}}{\sqrt{L_x}} 
 N(y) 
\begin{pmatrix}
 \sin \left( k_y y - \theta({\bf k}) \right) \cr
 \sin \left( k_y y \right) 
\end{pmatrix},
\label{eq:s-sol}
\end{align}
where $L_x$ is the length of the zigzag edge, and 
$N(y)=N \ne 0$ for $y \le 0$ and $N(y)=0$ otherwise. 
The value of $N$ is determined by the normalization condition.
We note that the standing wave of Eq.~(\ref{eq:zigwfsum})
reproduces the result of tight-binding lattice model
(see Appendix B of Ref.\onlinecite{sasaki09}).

First, we examine the Berry's phase of the standing wave.
For simplicity, 
let us consider the case $\theta({\bf k})= \pi/2$ for Eq.~(\ref{eq:s-sol}).
This corresponds to a normal incident process to the zigzag edge
[$k_x=0$ and $k_y>0$].
The amplitude for this process does not vanish as
\begin{align}
 \Psi^c_{{\rm K},k_y}({\bf r}) = \frac{N(y)}{\sqrt{L_x}}
\begin{pmatrix}
 -\cos \left( k_y y \right) \cr
 \sin \left( k_y y \right) 
\end{pmatrix}.
\end{align}
This standing wave includes a backward scattering amplitude, 
and the existence of the backward scattering amplitude indicates 
that the Berry' phase of the standing wave vanishes.
Indeed, since $\theta({\bf k}')= -\theta({\bf k})$, 
the Berry's phase of the reflected state is given by $+\pi$ as 
\begin{align}
 \int_0^{2\pi}  
 \langle \phi_{{\rm K},{\bf k'}}^c|
 \frac{1}{i}\frac{\partial}{\partial \theta}|
 \phi_{{\rm K},{\bf k'}}^c \rangle d\theta 
 = + \pi,
\end{align}
while that of the incident state is $-\pi$ [see Eq.~(\ref{eq:berry-})].
The Berry's phase of $-\pi$ for the incident state
is canceled by the Berry's phase of $\pi$ for the reflected state,
and hence the Berry's phase for the standing wave vanishes in total.
The trivial Berry's phase can be understood from Eq.~(\ref{eq:s-sol}),
in which the Bloch function of the standing wave is real~\cite{simon83}.

Next, we study the standing wave near the armchair edge.
The K and K$'$ points need to be considered simultaneously 
in the case of armchair edge 
since the armchair edge is the source of intervalley scattering.
Suppose that the wave vector of the incident wave 
measured from the K point is ${\bf k}=(k_x,k_y)$.
Thus, the wave vector of the elastically reflected wave measured from
the K$'$ point is given by ${\bf k'}=(-k_x,k_y)$, 
and the standing wave is written as
\begin{align}
 \Psi^c_{{\bf k}}({\bf r}) = 
 \Phi^c_{{\rm K},{\bf k}}({\bf r}) + e^{ia} \Phi^c_{{\rm K'},{\bf k'}}({\bf r}).
\end{align}
It can be shown that 
the Bloch state near the K$'$ point is given by
$\Phi^c_{{\rm K'},{\bf k'}}({\bf r}) = (e^{i{\bf k'}\cdot {\bf r}}/\sqrt{V})
 \phi^c_{{\rm K},{\bf k}}$~\cite{sasaki10-jpsj}.
Thus, we may write
\begin{align}
 \Psi^c_{{\bf k}}({\bf r}) 
 = \frac{e^{ik_y y}}{\sqrt{L_y}}N(x) \phi^c_{{\rm K},{\bf k}}
  \begin{pmatrix}
  e^{+ik_x x} \cr e^{i a} e^{-ik_x x} 
  \end{pmatrix},
\label{eq:armwf}
\end{align}
where $N(x)=N \ne 0$ for $x \le 0$ and $N(x)=0$ otherwise. 
The relative phase $e^{ia}$ 
can be determined by the boundary condition for armchair edge.
The value of $e^{ia}$ must be $-i$, which will be derived
elsewhere.~\cite{sasaki10-chiral}
Note that the Bloch functions for the K and K$'$ points
are the same. 
Thus, the Berry's phase of the standing wave near the armchair edge
is given by $-\pi$.
As a result, we can expect that 
the absence of backward scattering mechanism is valid
near the armchair edge.
Note, however, that 
the absence of backward scattering mechanism for the armchair edge
is not identical to that discussed so far for carbon nanotubes.
This is because the notion of a ``time-reversal pair'' of scattered
waves in the original argument~\cite{ando98} should be replaced to 
a true time-reversal pair of scattered waves near the armchair edge.
The time-reversal state of Eq.~(\ref{eq:armwf}) is given by~\cite{sasaki08ptps}
\begin{align}
 {\cal T}\Psi^c_{{\bf k}}({\bf r})= \frac{e^{-ik_y y}}{\sqrt{L_y}}
 N(x) (\phi^c_{{\rm K},{\bf k}})^*
 \begin{pmatrix}
  e^{-i a}  e^{+ik_x x} \cr 
  e^{-ik_x x}
 \end{pmatrix}.
\end{align}
A magnetic field breaks time-reversal symmetry, 
so that it invalidates the new absence of backward scattering
mechanism and leads to weak anti-localization.
It is also interesting to point out that 
the Berry's phase for the standing wave near the armchair edge
is robust against an ordinary mass term, 
while it is not robust against a topological mass term~\cite{haldane88}.

\begin{figure}[htbp]
 \begin{center}
  \includegraphics[scale=0.48]{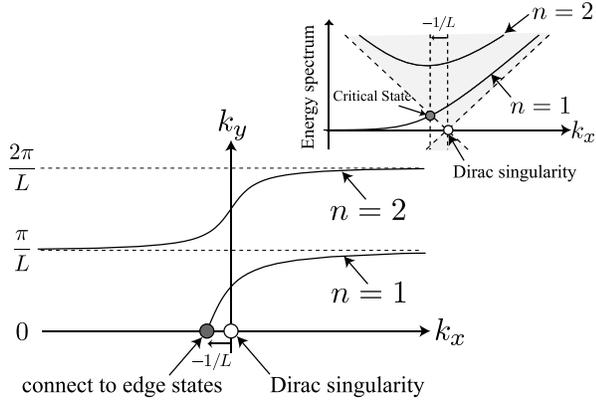}
 \end{center}
 \caption{The curves represent the solutions of parameters $k_x$ and
 $k_y$ which are allowed by the boundary condition for a zigzag
 nanoribbon. The Dirac singularity is not located on the curves, 
 which shows that there is no singularity at the parameter space of the
 standing waves. 
 The inset shows the corresponding spectral flow 
 in the energy dispersion relation. 
 }
 \label{fig:sflow}
\end{figure}

The appearance of a nontrivial Berry's phase 
is related to a singularity in parameter space~\cite{berry84,simon83}.
Here, we show the relationship between 
the singularity at the Dirac point (${\bf k}=0$)
in the momentum space and the standing wave.
For the zigzag edge, 
$k_x$ is a continuous variable, and therefore 
the spectrum with $k_y = 0$ may cross the Dirac singularity.
However, Eq.~(\ref{eq:s-sol}) shows that
the standing wave with $k_y =0$ does not exist,
that is $\Psi_{{\rm K},k_y=0}^c({\bf r}) = 0$.
Thus, the Dirac singularity seems to be separated from the spectrum
of the standing wave. 
To explore this point further, 
we consider a zigzag nanoribbon by introducing 
another zigzag edge at $y = -L$, in addition to the zigzag edge at $y=0$.
Suppose that the edge atoms at $y = -L$ are B-atoms, 
which imposes the boundary condition on the wave function at $y=-L$
as $\lim_{y=-L}\Psi_{{\rm K},{\bf k},{\rm A}}^c({\bf r}) = 0$.
This leads to the constraint equation for $(k_x,k_y)$,
\begin{align}
 k_y L + \theta({\bf k}) = n\pi,
 \label{eq:const}
\end{align}
where $n$ is an integer.
We note that this equation provides the solutions of 
$-k_x = k_y/\tan(k_yL)$ which was obtained by 
Brey and Fertig in Ref.~\onlinecite{brey06}
[the minus sign in front of $k_x$ is a matter of notation].
Note that $n$ must be a nonzero integer because 
the equation does not possess a solution when $n=0$.
In Fig.~\ref{fig:sflow},
we plot the solutions of Eq.~(\ref{eq:const}) for the cases 
$n =1$ and 2. 
The solution for $n >2$ is given by shifting the curve for $n=2$ 
by $\pi/L$,
and only the solution with $n=1$ shows an anomalous feature.
The curve with $n=1$ approaches a point on the $k_x$-axis.
This behavior can be checked by setting 
$\theta({\bf k}) = \pi + k_y/k_x$ in Eq.~(\ref{eq:const}) with $n=1$.
It is appropriate to refer to this state with $k_x = -1/L$ and $k_y = 0$ 
as the critical state~\cite{sasaki05prb}
because this state is neither the standing wave nor the edge
state~\cite{fujita96}.
The fact that no curve crosses the Dirac singularity
is consistent to the absence of a nontrivial Berry's phase for the
standing wave.
In other words, 
an electron can not approach the Dirac singularity
due to the presence of the zigzag edge.
As a result, 
an energy gap appears in the spectrum of the standing wave. 
The minimum energy gap is determined by the critical state as
$E_{\rm gap} = 2\hbar v_{\rm F}/L$ where $v_{\rm F}$ is the Fermi
velocity. 
In contrast to the boundary condition for the zigzag edge, 
the boundary condition for the armchair edge
does not forbid an electronic state 
cross the Dirac singularity point, 
and therefore the electron can pick up 
a nontrivial Berry's phase.

\begin{figure}[htbp]
 \begin{center}
  \includegraphics[scale=0.6]{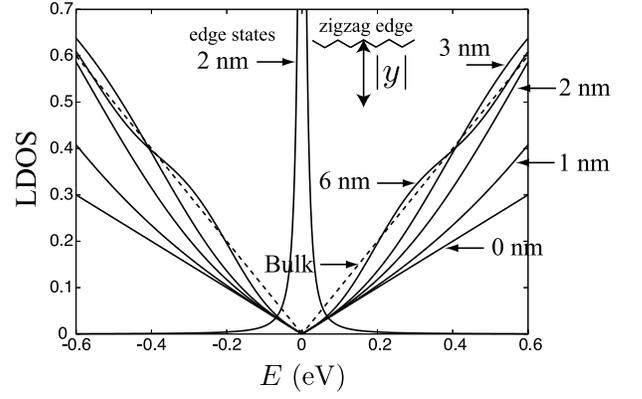}
 \end{center}
 \caption{
 The position dependence of the LDOS structure near the zigzag edge.
 The peak structure at $E=0$ is due to the edge states.
 }
 \label{fig:pspin}
\end{figure}

Here, we consider pseudospin of the standing wave
in order to examine the local density of states (LDOS) 
near the zigzag edge. 
The pseudospin for an eigenstate $\Psi(y)$ is defined by
the expectation value of Pauli matrices as 
$\langle \sigma_i \rangle \equiv \int \sigma_i(y) dy$
($i=x,y,z$) where $\sigma_i(y)$ is a pseudospin density
defined by 
$\sigma_i(y) \equiv \Psi^\dagger(y) \sigma_i  \Psi(y)$.
The boundary condition of Eq.~(\ref{eq:phiB0}) means that 
the pseudospin density is polarized 
into the positive $z$-axis locally near the zigzag edge, 
that is, $\sigma_z(0) > 0$ and $\sigma_x(0)=\sigma_y(0)=0$.
Actually, by putting $y \simeq 0$ into Eq.~(\ref{eq:s-sol}),
we see that the standing wave near the zigzag edge
has amplitudes only on A-atoms.
This polarization of the pseudospin causes an anomalous 
behavior to appear in the LDOS near the zigzag edge. 
To show this, let us first review the LDOS for 
a graphene without an edge. 
Assuming that electrons are non-interacting,
the bulk LDOS is given by
\begin{align}
 \rho(E) 
 = \frac{1}{2\pi} \frac{|E|}{(\hbar v_{\rm F})^2},
 \label{eq:ldos_bulk}
\end{align}
where $\rho(E)$ is proportional to $|E|$, 
which results from the Dirac cone spectrum.
The LDOS near the zigzag edge can be calculated 
by using the standing wave given in Eq.~(\ref{eq:s-sol}).
The LDOS has the form,
\begin{align}
 \rho_s(E,y)= \frac{1}{2\pi} \frac{|E|}{(\hbar v_{\rm F})^2} R(E,y),
\end{align}
where $R(E,y)$ is defined as
\begin{align}
 R(E,y) \equiv
 \frac{1}{\pi}\int^\pi_0 d\theta 
 \Psi_{{\rm K},{\bf k}}(y)^\dagger
 \Psi_{{\rm K},{\bf k}}(y).
 \label{eq:Rfac}
\end{align}
By performing the integral with respect to the angle $\theta$ 
in Eq.~(\ref{eq:Rfac}), we obtain the analytical result for $R(E,y)$ as
\begin{align}
 R(E,y) = 
 1-\frac{1}{2} \left\{ J_0\left(2\frac{|Ey|}{\hbar v_{\rm F}}\right)
 +J_2\left(2\frac{|Ey|}{\hbar v_{\rm F}} \right) \right\},
\label{eq:ldos}
\end{align}
where $J_\nu(x)$ is a Bessel function of order $\nu$.
The LDOS is symmetric with respect to $E=0$. 
The bulk LDOS in Eq.~(\ref{eq:ldos_bulk})
can be reproduced by setting 
$|y| \to \infty$ in Eq.~(\ref{eq:ldos})
since $J_\nu(\infty)=0$.
In Fig.~\ref{fig:pspin},
we plot the LDOS at $|y|=0$, 1, 2, 3, and 6[nm].
The slope of the LDOS depends on the distance from the zigzag edge.
At the edge, that is, at $|y|=0$, 
since $J_0(0)=1$ and $J_2(0)= 0$, 
we see that $R(E,0) =1/2$, and 
the slope of $\rho_s(E,0)$ is half of that of $\rho_s(E,\infty)$ [$=\rho(E)$].
This results from the fact that 
the amplitudes at one of two sublattice disappears 
near the zigzag edge (due to the pseudospin polarization), 
and only the half of the amplitude in the unit cell can contribute to
the LDOS.

In Fig.~\ref{fig:pspin}, 
in addition to the LDOS due to the standing wave, 
we plot the LDOS (at $|y|=2$[nm]) due to the edge
states~\cite{fujita96,nakada96}.
The edge states produce a peak at $E=0$ in the LDOS.
The LDOS of the edge states is calculated as follows.
The wave function of the edge state is given by~\cite{sasaki10-chiral}
\begin{align}
 \Psi_{{\rm K},k_x}({\bf r}) 
 = \frac{e^{ik_x x}}{\sqrt{L_x}}N(y)\sqrt{2|k_x|} e^{k_x |y|}
 \begin{pmatrix}
  1 \cr 0
 \end{pmatrix}.
\end{align}
Note that the edge states appear for $k_x < 0$ and $y<0$
(see Fig.~\ref{fig:sflow}).
Since the energy eigenvalue of the edge state vanishes,
the LDOS for the edge states is written as
\begin{align}
 \rho_e(E,y)
 = \frac{\delta}{\pi}
 \sum_{k_x < 0} \frac{\Psi_{{\rm K},k_x}({\bf r})^\dagger\Psi_{{\rm
 K},k_x}({\bf r})}{(E-0)^2+\delta^2},
\label{eq:rhoes}
\end{align}
where $\delta$ is a phenomenological parameter 
representing energy uncertainty of the edge states, 
for which we assume $\delta = 10$ meV.
Substituting $\sum_{k_x<0}$ with $(L_x/2\pi)\int_{-\infty}^0 dk_x$
in Eq.~(\ref{eq:rhoes}), and performing the integral for $k_x$, 
we obtain
\begin{align}
 \rho_e(E,y) = \frac{1}{2\pi^2}
 \frac{2\delta}{E^2+\delta^2}\frac{1}{4y^2}.
 \label{eq:dosed}
\end{align}
This result 
has been used in Fig.~\ref{fig:pspin}
to plot the LDOS of the edge states.
Note that $\rho_e(E,y)$ decreases as $\sim y^{-2}$, 
which is a slowly decreasing function 
compared with the exponential decaying wave function of the edge state.
It is interesting to note that $\rho_e(E,y)\sim y^{-2}$
can be derived from
the fact that $\int_0^{E_c} \{\rho_s(E,y)+\rho_e(E,y) \} dE$ does
not depend on $y$ for a large value of $E_c$.
This is a sum rule for the total LDOS consisting of the edge states and
the standing waves, 
in which the density above (below) the Fermi energy at $E=0$ must be
position independent.
This argument also shows that the singular behavior of the LDOS
for the edge states at $y=0$ is a consequence of 
the wave function of the standing wave.



By examining the standing wave near the graphene edge,
we have seen that 
the armchair edge preserves a non-trivial Berry's phase 
as in the bulk of graphene,
on the other hand, such a non-trivial phase is absent 
for the standing wave near the zigzag edge. 
Here, we consider the indication of this result 
with respect to the transport in graphene nanoribbons. 
In the argument for the absence of the backward scattering 
given in the introduction (see Fig.~\ref{fig:abs}), 
we find, by replacing $\Phi_{\bf k}$ with the standing wave $\Psi_{\bf k}$, that 
the backscattering amplitude, $\Psi_{\rm bs}=\left[1+U(2\pi)\right]\Psi_{\bf -k}$,
is enhanced for zigzag nanoribbons (due to $U(2\pi)=1$),
while it still vanishes for armchair nanoribbons (due to $U(2\pi)=-1$). 
Thus, we can expect that the conductance of a zigzag nanoribbon 
is smaller than that of an armchair nanoribbon, 
if we assume only scatters with long-range potential. 
The simulations performed by Areshkin {\it et al.}~\cite{areshkin07}
and Yamamoto {\it et al.}~\cite{yamamoto09}
show indeed that long-range potentials are ineffective in causing
backscattering in a perfect armchair nanoribbon. 
For narrow metallic armchair nanoribbons, 
the linear energy dispersion 
originating from the Dirac point provides a nearly
perfect conduction, and 
this nearly perfect conduction might relate to 
the nontrivial Berry's phase since the linear energy dispersion
picks up the Dirac singularity.
Note, however, that the standing wave for the armchair edge 
is a short-length intervalley mode, that is, the electronic wave length
of the standing wave is order of the lattice constant.
As a result, the wave function can be very sensitive 
to a short-range scattering potential, such as irregular edges~\cite{areshkin07} 
and lattice vacancies~\cite{lilu08}.
An armchair nanoribbon has both advantages (non-trivial Berry's phase)
and disadvantages (intervalley mode) for the transport 
as well as that a zigzag nanoribbon has both advantages (intravalley
mode) and disadvantages (trivial Berry's phase). 
Therefore, in a realistic situation, it is reasonable to consider
that a nanoribbon does not exhibit an electronic conduction 
comparable to a metallic carbon nanotube.

It seems that recent tight-binding numerical studies show that 
the conductance of zigzag nanoribbons is more robust than that of
armchair nanoribbons 
against edge disorders.~\cite{areshkin07,lilu08,cresti09,cresti09prb} 
In the numerical studies, 
the transport given by the electrons which are located very close
to the Fermi level is concerned.
For zigzag nanoribbons, the propagating
modes in each valley contain a single one-way excess channel 
(the states on the curve of $n=1$ which are close to the critical state 
shown in Fig.~\ref{fig:sflow}).
This feature causes a perfectly conducting channel to appear 
in the disordered system for the case that 
impurities do not give rise to intervalley
scattering~\cite{wakabayashi07,wakabayashi09njp}.
This feature also seems to cause a robust conducting channel to appear
in the edge disordered system.~\cite{areshkin07,lilu08,cresti09,cresti09prb} 
For our discussion of the backscattering in zigzag nanoribbons, 
we assume that 
the initial state with momentum $k_x$ accompanies the corresponding final
state with $-k_x$.
This condition is not satisfied for the single one-way excess channel,
for which our discussion for the Berry's phase is useless.
The single one-way excess channel plays a dominant role in zigzag
nanoribbons of the widths up to several nanometers for a limited energy
window.~\cite{areshkin07}
Our result for the Berry's phase may be useful in discussing transport 
of zigzag nanoribbons of the widths more than 10 nm,
for which we have several number of channels having the final state with
$-k_x$.

Our result may have a relationship to the formation of a transport
gap observed for graphene
nanoribbons.~\cite{stampfer09,han10,gallagher10}  
Note that the standing wave near a realistic rough edge 
can be described as the sum of
the standing waves for zigzag and armchair edges.~\cite{sasaki10-jpsj}
Since the standing wave for zigzag edge is a long-length intravalley
mode and that for armchair edge is a short-length intervalley one, 
their characters do not interfere with each other and 
can coexist in the standing wave near a rough edge.
This is also indicative of that 
a small portion of zigzag edge in a rough edge 
eventually governs the long-length transport behavior of a nanoribbon, 
as is numerically simulated by several authors.~\cite{evaldsson08,mucciolo09}
Indeed, Gallagher {\it et al.}~\cite{gallagher10} 
and Han {\it et al.}~\cite{han10}
observed that the source-drain gap
has a strong dependence on ribbon's length.
It is also important to recognize that 
there are two distinct localized states; 
the edge states and 
localized states which consist of the standing waves.
Since the localization length of the former 
is on the order of lattice constant,
it is difficult to consider that 
the transport behavior of nanoribbons 
of the widths larger than $10$ nm
is dominated by the edge states.~\cite{martin09}
In contrast, the latter is caused by the scatters, 
such as a potential disorder created by charge impurities, 
giving rise to backward scattering.
Since the latter consists of the standing waves, 
there is a possibility that 
the localization length is the order of the width of a ribbon.
Han {\it et al.}~\cite{han10} 
measured electron transport 
in lithographically fabricated 
nanoribbons of widths $20 < W < 120$ nm
with rough edge on the order of nanometers,
and confirmed that the size of a transport gap
is inversely proportional to the ribbon width.

In conclusion, 
we have shown that 
the Berry's phase for the standing wave near the zigzag edge is
trivial. 
The momentum parameter space for the standing wave near the zigzag edge 
does not include the Dirac singularity,
which is essential to the absence of a non-trivial Berry's phase.
As a result, the absence of backward scattering mechanism that works
well in a metallic carbon nanotube can not be used for a zigzag
nanoribbon. 
The absence of the Dirac singularity in the momentum space 
of the standing wave results in the peak in the LDOS 
due to the edge states. 
An observation of the LDOS peak near the zigzag edge is a direct
evidence of the absence of the Dirac singularity in the parameter space
of the standing wave. 
In contrast, a non-trivial Berry's phase survives the reflection by the
armchair edge.
For the standing wave near the armchair edge, 
we obtain the new absence of backward scattering mechanism in which 
a real time-reversal pair of backward scattered waves 
cancels with each other.
This absence of the backward scattering mechanism for the armchair edge 
is not robust against a short-range scattering potential
since the standing wave is a short-length intervalley mode.
An armchair nanoribbon has both advantage and disadvantage for the
transport as well as a zigzag nanoribbon.

\section*{Acknowledgment}

This work is financially supported by 
a Grant-in-Aid for Specially Promoted Research
(No.~20001006) from 
the Ministry of Education, Culture, Sports, Science
and Technology.

\bibliographystyle{apsrev}

\end{document}